\documentclass[journal]{IEEEtran}
\usepackage{graphicx}
\usepackage{color}
\usepackage{fancyhdr}
\usepackage[cmex10]{amsmath}
\usepackage{mdwmath}
\usepackage{amssymb}
\usepackage{textcomp}
\usepackage{times}
\usepackage{multicol}
\usepackage{float}
\usepackage[numbers,sort&compress]{natbib}

\usepackage{algorithm} 
\usepackage{algorithmic} 

\ifCLASSINFOpdf
\else
\fi
\begin{document}
%
\title{Wireless Communication System with RF-based Energy Harvesting: From Information Theory to Green System}
%
%



\author{Tao~Li,~Pingyi~Fan,~\IEEEmembership{Senior~Member,~IEEE}, Khaled Ben Letaief, ~\IEEEmembership{Fellow,~IEEE}
\thanks{T. Li and P. Fan are with the Tsinghua National Laboratory for Information Science and Technology, and Dept. of Electrical Engineering, Tsinghua University, Beijing, P. R. China., Emails: litao12@mails.tsinghua.edu.cn; fpy@tsinghua.edu.cn.

K. B. Letaief is with Dept. Electronic and Computer Engineering, Hong Kong University of Science and Technology, Email: eekhaled@ece.ust.hk.}}

\maketitle
\thispagestyle{empty}

\begin{abstract}
Harvesting energy from ambient environment is a new promising solution to free electronic devices from electric wire or limited-lifetime battery, which may find very significant applications in sensor networks and body-area networks. This paper mainly investigate the fundamental limits of information transmission in wireless communication system with RF-based energy harvesting, in which a master node acts not only as an information source but also an energy source for child node while only information is transmitted back from child to master node. Three typical structures: optimum receiver, orthogonal receiver and power splitting receiver are considered where two way information transmission between two nodes under an unique external power supply constraint at master node are jointly investigated in the viewpoint of systemic level. We explicitly characterize the achievable capacity-rate region and also discuss the effect of signal processing power consumption at child node. The optimal transmission strategy corresponding to the most energy-efficient status, namely the point on the boundary of achievable capacity-rate region, is derived with help of conditional capacity function. Simulation confirms the substantial gains of employing optimal transmission strategy and optimum receiver structure. Besides, a typical application on minimizing required transmit power to green system is presented.


\end{abstract}


%
\IEEEpeerreviewmaketitle

\section{Introduction}
%
%
%
%

\IEEEPARstart{H}{arvesting} energy from ambient environment is a promising solution to energy-constrained electronic devices, which are usually supported by battery with limited lifetime. For some special application scenarios, replacing battery is too expensive or even impossible to do, such as sensor network working under toxic environment and body-area network placed inside of human body,
where energy harvesting is a meaningful alternative technology. Excepting some kinds of renewable energy, such as solar and wind, wireless radio frequency (RF) signal also can be utilized as an important source for energy harvesting. Compared with other kinds of sources, RF-based energy harvesting, also called as wireless energy transfer, has some unique advantages. Since it is an active energy supply way, RF-base energy harvesting can provide more reliable energy flow to guarantee the performance of system. Thus, this paper concentrates on the energy harvesting system based on RF signal.

In fact, wireless energy transfer has been a topic of interest from the early 20th century until today, which was firstly proposed by Nikola Tesla \cite{Tesla_1}. A prototype system was even built to realize this goal in Canada at that time \cite{Garnica_2}. Due to the advance of high power vacuum tube and antenna technology, long distance wireless energy transfer became possible by microwave signal in 1960s, even between satellite and earth station \cite{Brown_3}. Recently, Andr¨¦ Kurs et al discussed wireless power transfer via strongly coupled magnetic resonances, in which it indicated that energy efficiency can reach $40\%$ within $2$ meters \cite{Kurs_4}. The work in \cite{Zhang_5} discussed the effect of relay on the efficiency and coverage of wireless energy transfer.

Wireless communication, another important application of electromagnetic signal based on Shannon theory \cite{Shannon_6}, has achieved great success in past several decades. Though both wireless information transmission and wireless energy transfer are built upon electromagnetic theory, they are always considered separately by electronic engineer and electrical engineer, respectively. It is intuitive that some performance gain may be obtained if they are considered jointly.

This problem was firstly investigated by Varshney in \cite{Varshney_7}, in which the optimal input signal is given when energy and information are transferred simultaneously. The work of Grover and Sahai in \cite{Grover_8} extended these results to frequency selective channel. In fact, energy and information transfer simultaneously has already been applied in some kind of practical systems, such as power line communication in wired system \cite{Meng_9} and radio frequency identification (RFID) in wireless system \cite{Baude_10}. Considering the fact that energy cannot be harvested from received signal after the information has been decoded, a novel energy harvesting receiver was proposed in \cite{Zhou_11} and a power splitting scheme was discussed in \cite{Liu_12}. The work in \cite{Huang_13,Zhang_14} discussed the performance of relay network and multiple-input-multiple-output (MIMO) system with energy harvesting. The work in \cite{Huang_15,Huang_16} discussed the application of energy harvesting in cellular networks.

Energy managing in energy harvesting system is another key problem that should be taken into account, since value of harvested energy is usually time-varying while traditional power supply is stable and invariant \cite{Ozel_17}-\cite{Ozel_24}. Ozel and Ulukus analyzed the achievable AWGN capacity under stochastic energy harvesting in \cite{Ozel_17,Ozel_18}, and an energy managing scheme called "store-and-transmit" was proposed to achieve the capacity. The optimal packet scheduling was discussed in terms of maximizing throughput in \cite{Yang_21,Yang_22}. The work in \cite{Tutuncuoglu_23} discussed the optimal transmission policies when the energy storage capacity was limited. And the work in \cite{Ozel_24} reconsidered this problem in fading channel scenarios.

According to above considerations, this paper investigates the fundamental limits of information transmission in wireless communication system with RF-based energy harvesting, which consists of a master node and a child node. It is assumed that master node acts not only as an information source but also an energy source for child node by wireless energy transfer while only information is transmitted back from child node to master node. That is to say, the whole system is powered by an unique external power supply at master node and information is exchanged between two nodes, which is a typical system structure in many scenarios, such as sensor networks and body-area networks.

From a perspective of information transmission, the ultimate goal of energy harvesting at child node is to support the information transmission from child node to master node. Thus, two output variables of the system, namely two way information rates between master node and child node, need to be maximized under limited transmit power at master node. Obviously, there exists a tradeoff relationship between them. \emph{Achievable capacity-rate region} is proposed by this paper to characterize the tradeoff relationship in this case. To the best of our knowledge, it hasn't been considered in the existing literature. It's worth noting that the system model in this paper is different with these in \cite{Zhang_14,Ju_32,Chen_33} since two way information transmission under an unique external power supply at master node are jointly considered by this paper from a perspective of system level. Namely, the work in \cite{Zhang_14} focused on the problem for transferring energy and information simultaneously from master to child node, and the works in \cite{Ju_32,Chen_33} considered wireless energy transfer in the downlink and information transmission in the uplink.

Part of this work has been published in \cite{Li_32}, which analyzed the transmission performance in optimum receiver system and orthogonal receiver system. This paper extends these results to typical power splitting receiver system. In a time division duplex system, the optimal transmission strategies under three different receiver structures that correspond to the points located on the boundary of \emph{achievable capacity-rate region} are also derived in terms of green system, which are significantly meaningful for practical system design since the energy harvested from RF signal is so rare and precious due to natural path loss of electromagnetic signal. For the same reason, power consumption by signal processing at energy harvesting node is also taken into account based on the results in \cite{Zhang_25,Zhang_26,Grover_27,Grover_28} throughout this paper, which can be modeled as a constant power expenditure once transceiver sets up \cite{Orhan_31}. Lastly, a typical application in terms of minimizing required transmit power, namely green system, is introduced to validate the proposed results.

The rest of this paper is organized as follows. In Section II, system model and wireless energy/information receiver are presented. Simultaneous wireless information and energy transferring from master to child node are introduced under three different receiver structures in Section III, while information transmission and energy managing strategy at child node are discussed in Section IV.
Then, we investigate the tradeoff relationship between two way information rates in Section V, where some simulation results are also given. At last, a typical application of the system proposed by this paper for body-area network is introduced in Section VI.

\section{Preliminary}

\subsection{System Structure}

\begin{figure}[!t]
\centering
\includegraphics[width=3.2 in]{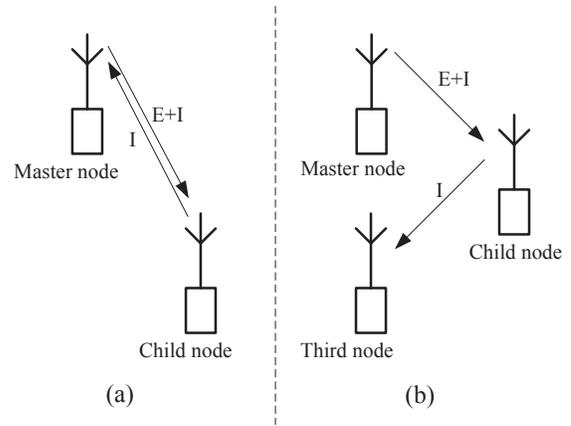}
\caption{Two instances of wireless communication system with RF-based energy harvesting, in which E denotes energy flow and I denotes information flow.}
\label{fig:system_framework}
\end{figure}

As is stated before, this paper concentrates on wireless communication system with RF-based energy harvesting, such as data acquisition system and sensor network, and attempts to obtain the limits of information transmission performance under unique external power supply.
Fig. \ref{fig:system_framework} gives two simple instances of elementary units for this kind of system. Taking point-to-point communication system shown in Fig. \ref{fig:system_framework}(a) for example, two nodes consist of this system, which are called as master node and child node, respectively. Child node has the ability to harvest energy from received RF signal transmitted by master node. Thus, information is exchanged between them while the master node also acts as an external energy source for child node by wireless electromagnetic transmission. The link from master node to child node is defined as downlink, in which both energy and information will be transferred simultaneously. And the link from child to master node is defined as uplink, in which only information is transferred by radio carrier.
Besides, it's worth noting that the terms of \emph{downlink} and \emph{uplink} proposed here is just for description convenience, not suggesting that a cellular scenario is considered in this paper.

It is assumed that the channel environment is additive white Gaussian noise and there is only one antenna equipped at each node. In order to realize information transmission powered by harvested energy, child node should have the functions of energy receiver and information transceiver at the same time, which will be presented in detail in the sequel.
Besides, Fig. \ref{fig:system_framework}(b) gives another typical example for RF-based energy harvesting system, where information is transmitted from child node to a third node in the uplink phase. Though this paper mainly focuses on the RF-based energy harvesting system described in Fig. \ref{fig:system_framework}(a), similar conclusions can be obtained for the system illustrated in Fig. \ref{fig:system_framework}(b).


\subsection{Wireless Energy Transfer Receiver}

\begin{figure}[!t]
\centering
\includegraphics[width=3.3 in]{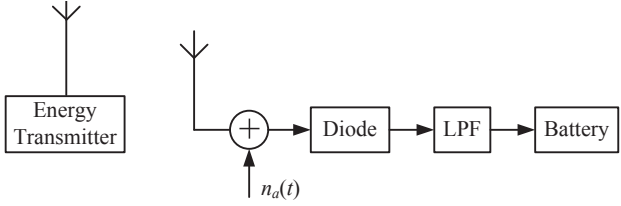}
\caption{The system framework of a simple energy harvesting system.}
\label{fig:energy_receiver_framework}
\end{figure}

Fig. \ref{fig:energy_receiver_framework} illustrates a simple example of wireless energy transfer system. Firstly, the energy transmitter transforms electric energy into electromagnetic energy, which is transmitted into free space by the antenna. Through free space, the RF signal is received by the antenna at energy receiver. Then, it is transformed from alternating current (AC) into direct current (DC) by diode and low-pass filter (LPF). After that, the energy can be stored into the battery at child node for future usages, such as information transmission and signal processing.

Assuming the transmit signal is $x(t)$ with $\mathbb{E}[|x(t)|^2]=1$, where $\mathbb{E}[\cdot]$ denotes statistical expectation operation. $x(t)$ is a narrow-band signal and the average transmit power is denoted as $P_0$. The frequency bandwidth of $x(t)$ is $B$ and the center of carrier frequency is $f_0$, $f_0\gg B$. Under these assumptions above, the received signal $y(t)$ can be expressed as
\begin{equation}\label{equ:expression of receiving signal}
  y(t)=\sqrt{\frac{GhP_0}{d^\alpha}}x(t)e^{j(2\pi f_0 t+\theta)}+n_a(t)
\end{equation}
where $G$ denotes a constant power gain generated by transmitter and receiver antennas, $h$ denotes power gain coefficient of channel, $d$ denotes the distance between transmitter and receiver, and $\alpha$ denotes the path loss exponent ($2\leq \alpha \leq 4$). $n_a(t)$ is additive circularly symmetric complex Gaussian noise with zero mean and power spectrum density $N_0$.

It is assumed that input signal $x(t)$ is independent with additive noise signal $n_a(t)$. When the average transmit power constraint is $P_0$, based on the law of conservation of energy and the expression of received signal in Eqn. (\ref{equ:expression of receiving signal}), the maximal average power that can be harvested is
\begin{equation}\label{equ:model for energy receiver}
  P_{h,max}=\eta \mathbb{E}[|y(t)|^2] = \eta \frac{Gh}{d^\alpha} P_0
\end{equation}
where $\eta$ denotes the conversation efficiency factor of energy harvesting receiver. Besides, the power contributed by noise signal is neglected compared with the received signal power.

It can be observed from Eqn. (\ref{equ:model for energy receiver}) that the value of energy transfer capacity has nothing to do with the realization of stochastic signal $x(t)$. Thus, Gaussian signal, which is often used as input signal for information transfer, also can be used to achieve energy transfer capacity. Without loss of generality, we assume a whole signal block time is $T=1$ in the sequel so that the terms of energy and power are interchangeable for description convenience.

\subsection{Wireless Information Transmission Receiver}

\begin{figure}[!t]
\centering
\includegraphics[width=3.0 in]{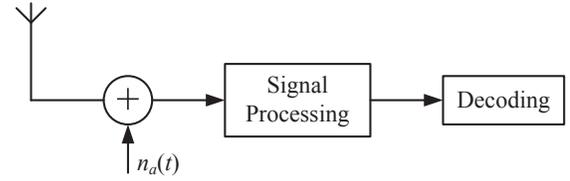}
\caption{The system framework of a typical information receiver.}
\label{fig:information_receiver_framework}
\end{figure}

Fig. \ref{fig:information_receiver_framework} illustrates the system structure of a simple information receiver, which consists of RF antenna, signal processing and decoding components. It is assumed that the system parameters are just the same as these introduced in the previous subsection. Based on the Eqn. (\ref{equ:expression of receiving signal}), the channel capacity for information transmission can be expressed as follows, which can be achieved by zero-mean, circularly symmetric complex Gaussian inputs.
\begin{equation}\label{equ:channel information capacity}
  C=B \cdot \log(1+\frac{GhP_0}{d^\alpha \sigma_0^2 })
\end{equation}
where $B$ denotes the available frequency bandwidth, and $\sigma_0^2$ denotes the additive noise power $\sigma_0^2=B N_0$. The base of logarithm operation is $2$ in this paper without any special declaration.

According to above results, \emph{Lemma 2.1} can be concluded, which will play an important role for simultaneous energy and information transfer in the downlink phase.

\emph{\textbf{Lemma 2.1:} In additive white Gaussian noise channel, zero mean and cyclic symmetric complex Gaussian signal is optimum for both wireless energy transfer and information transmission.}

\section{Simultaneous Energy and Information Transfer in The Downlink}

As is stated in Section II, energy and information are transferred from master node to child node simultaneously by wireless electromagnetic signal in the downlink phase, the performance of which can be characterized by achievable capacity-energy region. Since the structure of receiver has natural effect upon transmission performance, several typical examples will be reviewed firstly in this section, which contain optimum receiver system, orthogonal receiver system and power splitting receiver system. Then, some simulation and comments for simultaneous energy and information transfer will be given in terms of achievable capacity-energy region.

\subsection{Optimum Receiver System}

Simultaneous energy and information transfer was firstly proposed in \cite{Varshney_7}, which is performed with assumption that energy can be harvested after information has been decoded from the signal. That is to say, the receiver can obtain both energy and information from the same received signal without loss. It can be regarded as the upper case for energy and information simultaneous transfer. For description convenience, we refer the system in \cite{Varshney_7} as \emph{Optimum Receiver System}.

As the energy is so rare and precious at child node, signal processing power consumption should also be taken into account, denoted as $P_p$, which can be modeled as a constant power expenditure once transceiver sets up \cite{Orhan_31}. It is assumed that $R_d$ denotes the average information transmission rate in the downlink, and $P_s$ denotes the average energy that can be saved into the battery at child node for future usage.
Similar to the work in \cite{Zhou_11,Liu_12}, achievable capacity-energy region is used as a metric to evaluate the system performance of simultaneous energy and information transfer, the definition of which is given as follows.

\emph{\textbf{Definition 3.1:} The achievable capacity-energy region $\mathbb{T}_{\{R_d-P_s\}}$ is defined as the set that contains all possible achievable information transmission rate and stored energy pair $(R_d, P_s)$ which satisfy average transmit power constraint at master node.}

Thus, with zero mean circularly symmetric complex Gaussian input signal, the achievable capacity-energy region in \emph{Optimum Receiver System} can be expressed as
\begin{equation}\label{equ:achievable capacity-energy region for R-d and P-s}
\begin{split}
  \mathbb{T}_{\{R_d-P_s\}} \,\, = &\,\,\, \{(R_d,P_s)\,|\,0 \leq P_s+P_p \leq \eta \tfrac{Gh}{d^\alpha} P_0,  \\
      & \,\,\,\,\,\,\,\,\,\,\,\,\,\,\, 0 \,\leq \,R_d\, \leq\, B\log(1+\tfrac{Gh P_0}{d^\alpha \sigma_0^2})\}
\end{split}
\end{equation}

\subsection{Orthogonal Receiver System}

Since the assumption in \cite{Varshney_7} may be not allowed by practical physical circuit, some practical receiver structures are proposed in the literatures \cite{Zhang_14,Huang_16,Ju_32}. Generally, from a perspective of practical physical circuit, in order to achieve the goal of transferring energy and information simultaneously without interference, the signal intended for energy sub-receiver and information sub-receiver should be orthogonal in certain domain, such as time domain and frequency domain. Without loss of generality, it is assumed that the ratio of power at master node allocated for the goal of energy transfer is $\rho$ ($0 \leq \rho \leq 1$) in the sequel, and the other part is for information transmission. Thus, we call this kind of system as \emph{Orthogonal Receiver System}.

Taking time-orthogonal receiver system for example, similar results can be extended for other orthogonal receiver system easily, the ratio of time slots allocated for energy harvesting with respect to the whole transmission period is denoted as $\tau$ ($0\leq \tau \leq 1$). Namely, in the first $\tau T$ time slots, where $T$ denotes the length of transmission period, the received signal energy is used for energy harvesting, otherwise the energy is used for information decoding. The energy flow and information flow in this case are independent with each other in terms of time. 

\emph{\textbf{Proposition 3.1:}When the ratio of transmit power allocated for energy sub-receiver is $\rho$ in time-orthogonal receiver system, it can be demonstrated that the achievable capacity-energy region under optimal time ratio allocation strategy is}
\begin{equation}\label{equ:achievable capacity-energy region for R-d and P-s in TPSS}
\begin{split}
  \mathbb{T}_{\{R_d-P_s\}} =\bigcup \limits_{0\leq \rho \leq 1}&\{(R_d,P_s)\,|\,0 \leq P_s+P_p \leq \eta \tfrac{\rho Gh}{d^\alpha}P_0,  \\
                        & 0\, \leq \, R_d \,\leq \, B\log(1+\tfrac{Gh (1-\rho)P_0}{d^\alpha \sigma_0^2})\}
\end{split}
\end{equation}
\begin{proof}
 See Appendix A.
\end{proof}

\subsection{Power Splitting Receiver System}

\begin{figure}[!t]
\centering
\includegraphics[width=3.2 in]{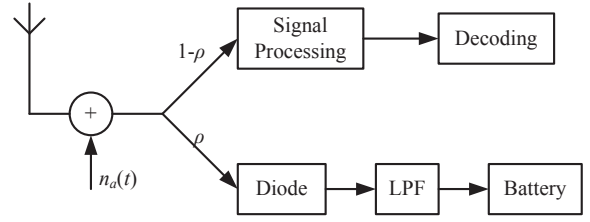}
\caption{The structure of a power splitting receiver system.}
\label{fig:power_splitting_framework}
\end{figure}

Excepting optimum receiver system and orthogonal receiver system, a novel receiver structure called as power splitting receiver system is also proposed for energy and information simultaneous transfer in \cite{Zhou_11,Liu_12,Huang_13,Zhang_14}, which is illustrated in Fig. \ref{fig:power_splitting_framework}. 
The transmit power for both energy sub-receiver and information sub-receiver occupies overall time and frequency resource in wireless channel. Then, it is separated into two sub-flows at the receiver based on the value of $\rho$, which is prior known by transmitter and receiver. The splitting point is located between RF component and signal processing component.
Without loss of generality, it is assumed the power splitting process at receiver is ideal so that there is no energy loss and there is no new noise introduced by it. Under this case, the overall additive noise $\sigma_0^2$ comes from two aspects: RF component and signal processing component, which are denoted as $\sigma_a^2$ and $\sigma_p^2$, respectively, where $\sigma_a^2+\sigma_p^2=\sigma_0^2$. Thus, the average information rate in this case can be expressed as
\begin{equation}\label{equ:information capacity in PSRS}
  R_d \leq B \log(1+\frac{Gh}{d^\alpha} \cdot \frac{(1-\rho) P_0}{(1-\rho) \sigma_a^2+\sigma_p^2})
\end{equation}

Similarly, the corresponding stored energy power $P_s$ meets the following inequality constraint
\begin{equation}\label{equ:energy capacity in PSRS}
  P_s+P_p\leq P_h \leq \eta \frac{Gh}{d^\alpha} \rho P_0
\end{equation}

Combining the results in (\ref{equ:information capacity in PSRS}-\ref{equ:energy capacity in PSRS}), the corresponding achievable capacity-energy region is
\begin{equation}\label{equ:achievable capacity-energy region for R-d and P-s in PSRS}
\begin{split}
  \mathbb{T}_{\{R_d-P_s\}}=&\bigcup\limits_{0 \leq \rho \leq 1} \{(R_d,P_s)\,|\,0\leq P_s+P_p \leq \eta \tfrac{Gh}{d^\alpha} \rho P_0, \\
           &\,\,\,\,\, 0 \leq  R_d \leq  B \log(1+\tfrac{Gh}{d^\alpha} \cdot \tfrac{(1-\rho) P_0}{(1-\rho) \sigma_a^2+\sigma_p^2})\,\,\}
\end{split}
\end{equation}

It can be observed from Eqn. (\ref{equ:achievable capacity-energy region for R-d and P-s in PSRS}) that the power splitting receiver system will degrade to orthogonal system if $\sigma_a^2=0$. And it can be considered as the optimum receiver system if $\sigma_p^2=0$ and $\rho$ is equal to $1$. In summary, optimum receiver system and orthogonal receiver system can be regarded as two extreme cases for power splitting receiver system.

\subsection{Performance Analysis and Discussion}

The expressions of achievable capacity-energy region in three different systems have been given in Eqn. (\ref{equ:achievable capacity-energy region for R-d and P-s}), (\ref{equ:achievable capacity-energy region for R-d and P-s in TPSS}) and (\ref{equ:achievable capacity-energy region for R-d and P-s in PSRS}). However, the boundary of region $\mathbb{T}_{\{R_d-P_s\}}$ is more meaningful for system design since it reflects the tradeoff relationship between $R_d$ and $P_s$, and characterizes the fundamental limits of system performance. Fixed the stored energy power $P_s$, the maximum achievable value of $R_d$ can be used to describe the point located on the boundary of capacity-rate region. Thus, we give the concept of capacity-energy function.

\emph{\textbf{Definition 3.2:} Capacity-energy function $C(P_s)$ is defined as maximum conditional information transmit capacity when average transmit power and stored energy are constrained to $P_0$ and $P_s$. Namely, $C(P_s)$ can be expressed as}
\begin{equation}\label{equ:expression of definition 3.2}
  C(P_s)=sup \{R_d:(R_d,P_s) \text{ is achievable}\}.
\end{equation}

Based on tradeoff relationship, when $R_d$ is zero, $P_s$ can achieve the maximal value, which is
\begin{equation}\label{equ:maximal energy in downlink}
  P_{s,max}=\eta \tfrac{Gh}{d^\alpha} P_0-P_p
\end{equation}

According to the results in Eqn. (\ref{equ:expression of definition 3.2})-(\ref{equ:maximal energy in downlink}), the achievable capacity-energy region $\mathbb{T}_{\{R_d-P_s\}}$ can be rewritten in terms of capacity-energy function $C(P_s)$.
\begin{equation}\label{equ:equivalent achievable energy-rate region}
  \mathbb{T}_{\{R_d-P_s\}}=\{(R_d,P_s)|0 \leq P_s \leq P_{s,max},0 \leq R_d \leq C(P_s)\}
\end{equation}

The residual problem is to get the expression of $C(P_s)$ corresponding to the Eqn. (\ref{equ:achievable capacity-energy region for R-d and P-s in TPSS}) and (\ref{equ:achievable capacity-energy region for R-d and P-s in PSRS}). In orthogonal receiver system, when stored energy is fixed to $P_s$ and average transmit power is $P_0$, the expression of $C(P_s)$ can be derived as a function of $P_s$ by eliminating $\rho$ as follows
\begin{equation}\label{equ:capacity-energy function in orthogonal receiver system}
   C(P_s) = B \cdot \log\{1+\tfrac{GhP_0-d^\alpha(P_s+P_p)/\eta}{d^\alpha \sigma_0^2} \}
\end{equation}

Similarly, the capacity-energy function in power splitting receiver system is
\begin{equation}\label{equequ:capacity-energy function in power splitting receiver system}
   C(P_s) =  B \log\{1+\tfrac{GhP_0}{d^\alpha} \cdot \tfrac{\eta GhP_0-d^\alpha(P_s+P_p)}{[\eta GhP_0-d^\alpha(P_s+P_p)]\cdot \sigma_a^2 +\sigma_p^2}\}
\end{equation}

For the system shown in Fig. \ref{fig:system_framework}(a), assuming the distance between master node and child node $d$ is $50 m$, the antenna power gain $G$ is $10 dB$, path loss component $\alpha$ is 2, channel power gain coefficient $h$ is $1$ and the frequency bandwidth $B$ is $1$ KHz. Without loss of generality, the power transformation efficiency is $\eta=1$ in this paper without specific declaration. The overall additive noise power $\sigma_a^2$ is $1 mW$. The basic power consumption for signal processing component $P_p$ is $5 mW$. The average transmit power at master node $P_0$ is $10 W$.

\begin{figure}[!t]
\centering
\includegraphics[width=3.2 in]{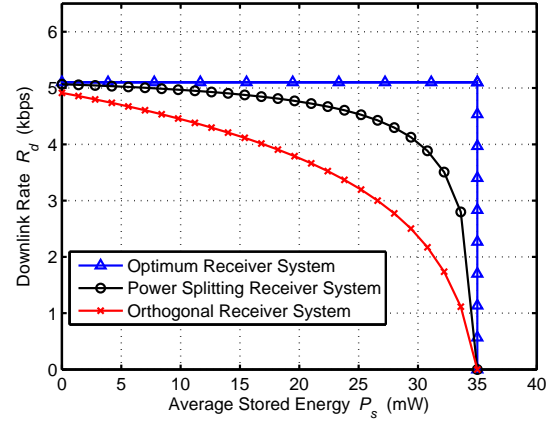}
\caption{The achievable capacity-energy region $\mathbb{T}_{\{R_d-P_s\}}$ for simultaneous information and energy transfer in the downlink phase, which contain three cases: orthogonal receiver system, optimum receiver system and power splitting receiver system with $\sigma_p^2/\sigma_a^2=1/4$.}
\label{fig:capacity_energy_region_three_case}
\end{figure}

Under these assumptions, Fig. \ref{fig:capacity_energy_region_three_case} illustrates three corresponding capacity-energy regions, which contain three cases: optimum receiver system, orthogonal receiver system and power splitting receiver system with $\sigma_p^2/\sigma_a^2=1/4$. As observed from Fig. \ref{fig:capacity_energy_region_three_case}, though signal processing power consumption at child node is also taken into account, the simulation result is consistent with that obtained in \cite{Zhang_14}. An obvious difference is that $R_{d,max}$ values in three different cases aren't the same due to $P_p$.
Besides, the optimum receiver system is best, and orthogonal receiver system can be viewed as a lower bound for simultaneous energy and information transferring.

\begin{figure}[!t]
\centering
\includegraphics[width=3.2 in]{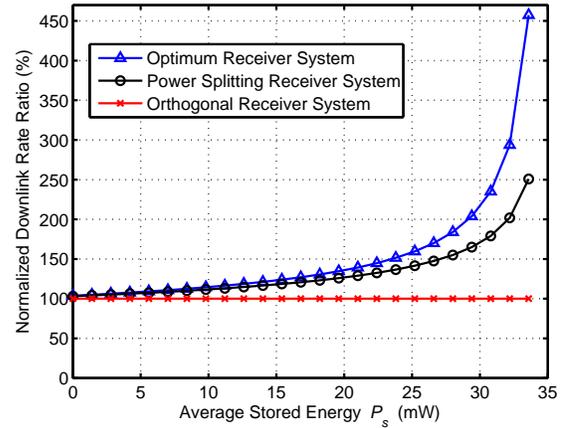}
\caption{The normalized maximum achievable downlink rate ratio as a function of average stored energy $P_s$ with respect to that in orthogonal receiver system when average transmit power at maser node is $10 W$.}
\label{fig:capacity_energy_region_three_case_normalized}
\end{figure}

For providing more insights, Fig. \ref{fig:capacity_energy_region_three_case_normalized} illustrates the normalized maximum achievable downlink rate ratio as a function of average stored energy $P_s$ with respect to that in orthogonal receiver system when average transmit power at maser node is constrained to $10 W$ (the point that $R_d=0$ is ignored). It can be seen that optimum receiver system and power splitting system apparently outperform the orthogonal receiver system, especially when $P_s$ is big. Since the energy flows for energy sub-receiver and information sub-receiver are independent with each other in orthogonal receiver system, it cannot benefit from the cooperation in simultaneous energy and information transfer process. The relative performance gain even can reach more than $100\%$ in some case.

\section{Information Transmission in the Uplink phase}

Information transmission in the uplink phase is powered by a stochastic arrival power source harvested in the downlink, which is different from traditional communication system. It is reasonable to assume that there is an energy storage at child node, such as battery, which can store energy without loss. According to the state of the art, the capacity of energy storage can be regarded as infinity in practical systems \cite{Ozel_17} compared with harvested energy.

Let $C$ denotes Shannon channel capacity when the transmitter is powered by constant power supply $P_s$. Since there exists causal relationship constraint between harvested energy and consumed energy in this energy harvesting system, the maximal achievable throughput $R_u$ is apparently less than $C$. However, the work in \cite{Ozel_17} has shown that $R_u$ can be asymptotic to the channel capacity $C$ if optimal energy managing scheme is employed by child node. \emph{Save-and-transmit strategy}, which is an optimal energy managing scheme, was proposed to achieve the maximal throughput. The corresponding conclusion is expressed as \emph{Lemma 4.1}.

\emph{\textbf{Lemma 4.1 \cite{Ozel_17}:} In AWGN channel, the maximal throughput udder i.i.d. random arrival power supply $P_s(n)$, where $\mathbb{E}[P_s(n)]=P_s$, is independent of the realizations of $P_s(n)$ and equal to the channel capacity $C$ only with average power constraint $P_s$:}
\begin{equation}\label{equ:result of Lemma 1}
  R_u = B \cdot \log (1+\tfrac{Gh P_s}{d^\alpha \sigma_0^2})
\end{equation}

The randomness of harvested energy and signal processing power consumption are jointly considered in this section. In a time slot based system, assuming the length of each time slot is $\Delta t$, the instantaneous stored power harvested from downlink and transmit power in the uplink during $n-th$ time slot are $P_s(n)$ and $P_t(n)$, respectively. $P_s(n)$ is usually identical and independent (i.i.d.) stochastic process with statistical mean $P_s$. Other system parameters are just the same as these in the downlink. Then the maximal achievable uplink data rate $R_u$ can be modeled as

\begin{subequations}\label{equ:optimal problem under random arrival energy}
\begin{align}
  R_u \,\,\, &= \lim \limits_{N \rightarrow +\infty} \{ \max \limits_{\{P_t(n)\}}  \frac{1}{N} \sum_{n=1}^{N} C(n) \}    \tag{15} \\
  s.t. \,\, & C(n)=B \log (1+\tfrac{Gh}{d^\alpha} \cdot \tfrac{P_t(n)}{\sigma_0^2}) \label{equ:optimal problem constraint under random arrival energy a} \\
            & \mathbb{E}[P_s(n)]=P_s, \label{equ:optimal problem constraint under random arrival energy b}  \\
            & \sum_{i=1}^{n} P_s(i)\Delta t+E_s(0) \geq \sum_{i=1}^{n} (P_t(i)+P_p)\Delta t, n=1,2...N  \label{equ:optimal problem constraint under random arrival energy c}
\end{align}
\end{subequations}
where $E_s(0)$ denotes the electric value in energy storage at $n=0$ time slot, $P_p$ denotes the signal processing power consumption and $C(n)$ denotes instantaneous transmission capacity at $n-th$ time slot. Eqn. (\ref{equ:optimal problem constraint under random arrival energy b}) indicates the average power constraint for stored energy derived in the previous section. The constraint in (\ref{equ:optimal problem constraint under random arrival energy c}) reflects causal relationship between harvested energy and consumed energy.

Based on the problem in (\ref{equ:optimal problem under random arrival energy}), since the convex property of objective function hasn't been changed by $P_p$, regardless the realization of $C(n)$, \emph{save-and-transmit strategy} is still the optimal power allocation strategy. Thus, the maximal achievable uplink information rate $R_u$ in this case can be summarized as the following proposition.

\emph{\textbf{Proposition 4.1:} In AWGN channel, if both transmit power and processing power are taken into account, the maximal achievable throughput $R_u$ under i.i.d. random arrival $P_s(n)$, where $\mathbb{E}[P_s(n)]=P_s$, is independent of the realizations of $P_s(n)$ and equal to channel capacity only with average power constraint $P_s$. That is to say, $R_u$ can be calculated as follows:}
\begin{equation}\label{equ:result of Proposition 1}
  R_u=B \cdot \log \{1+\tfrac{Gh \cdot (P_s-P_p)}{d^\alpha \sigma_0^2 } \}
\end{equation}
\begin{proof}
 See Appendix B.
\end{proof}

\section{The Joint Optimal Transmission Scheme for Downlink and Uplink}

This section investigates the limits of downlink and uplink information transmission from an information theoretical view. Though full-duplex strategy is optimal and there are some advances to apply it into practical system \cite{Jain_30}, it is very difficult to apply it into energy harvesting system due to the limit of signal processing capability at child node. Thus, time-division duplex (TDD) is employed in this section without any specific declaration.

\begin{figure}[!t]
\centering
\includegraphics[width=3.0 in]{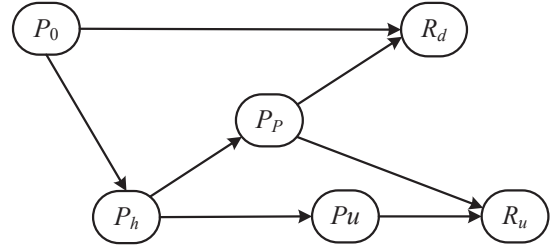}
\caption{The diagram of energy consumption flow in wireless communication system with RF-based energy harvesting}
\label{fig:power_consumption_schematic}
\end{figure}

Assuming that $P_u$ denotes transmit power at child node for uplink transmission, $R_d$ and $R_u$ denote average information rates of downlink and uplink, respectively. Considering power consumption from a systematic level, transmit power $P_0$ at master node is the unique external power supply for whole system. The energy flow based the types of usage is illustrated in Fig. \ref{fig:power_consumption_schematic}. It can be seen that the whole system is powered by an unique power supply $P_0$ at master node, two functional objectives $R_d$ and $R_u$ are the outputs of system. Obviously, there exists a tradeoff between $R_d$ and $R_u$, which will be investigated in optimum receiver system, orthogonal receiver system and power splitting receiver system in the sequel, respectively.

\subsection{Optimal Transmission Policy in Optimum Receiver System}

Firstly, in order to characterize the tradeoff relationship between $R_d$ and $R_u$, we give the definition of achievable capacity-rate region to be used as a metric for measuring system performance under average transmit power constraint when downlink and uplink are jointly considered.

\emph{\textbf{Definition 5.1:} Achievable capacity-rate region $\mathbb{T}_{\{R_d-R_u\}}$ denotes the set that contains all possible achievable instance $(R_d,R_u)$ which satisfy average power constraint at host node.}

It is assumed that time resource ratio allocated for downlink transmission with respect to a whole period is $\lambda$ ($0 \leq \lambda \leq 1$). Then, the achievable capacity-rate region in this case is
\begin{equation}\label{equ:the constraints for joint downlink and uplink transmission}
\begin{split}
 & \mathbb{T}_{\{R_d-R_u\}}=\bigcup \limits_{0 \leq \lambda \leq 1} \{(R_d,R_u)|0 \leq R_d \leq \lambda B \log(1+  \\
 & \,\,\,\,\, \tfrac{Gh P_0}{\lambda d^\alpha \sigma_0^2}), 0 \leq R_u \leq (1-\lambda) B \log(1+\tfrac{Gh P_u}{(1-\lambda) d^\alpha \sigma_0^2}) \}
\end{split}
\end{equation}
where the transmit power $P_u$ at child node in the uplink phase satisfies the following constraint.
\begin{equation}\label{equ:the constraint for transmit power at sensor node}
 P_p+P_u \leq \lambda \eta \tfrac{Gh}{\lambda d^\alpha}P_0
\end{equation}

Though the expression of $\mathbb{T}_{R_d-R_u}$ has been given in Eqn. (\ref{equ:the constraints for joint downlink and uplink transmission}), boundary curve around the region is more important for us since it reflects the fundamental limits of system performance and tradeoff relationship between $R_d$ and $R_u$. Thus, we give the definition of \emph{capacity-rate function}.

\emph{\textbf{Definition 5.2:} Capacity-rate function $C(R_d)$ denotes the maximal achievable uplink transmission rate on the condition of downlink information rate $R_d$, when the average transmit power at master node is $P_0$. It can be expressed as}
\begin{equation}\label{equ:expression of definition 5}
  C(R_d)=sup \{R_u:(R_d,R_u) \text{ is achievable}\}.
\end{equation}

According to \emph{Definition 5}, $C(R_u)$ can be modeled as the solution to the following problem.
\begin{subequations}\label{equ:optimal TDD for jointly considering uplink and downlink in Optimum system}
\begin{align}
 \max \limits_{\lambda} \,\,& R_u= (1-\lambda)B\log(1+\tfrac{Gh P_u/(1-\lambda)}{ d^\alpha \sigma_0^2})  \tag{20}\\
s.t.\,\,\,\,\, &   P_p + P_u \leq \lambda \eta \tfrac{Gh}{\lambda d^\alpha} P_0  ;\,\,\,\,\,\,\,\,\,\,\,\,\,\,\,\,\,\, \label{equ:optimal TDD constraint for jointly considering uplink and downlink in Optimum system a} \\
   & R_d \leq \lambda B \log(1+\tfrac{Gh P_0/\lambda}{ d^\alpha \sigma_0^2}); \label{equ:optimal TDD constraint for jointly considering uplink and downlink in Optimum system b} \\
   & R_d,R_u \geq 0. \label{equ:optimal TDD constraint for jointly considering uplink and downlink in Optimum system c}
\end{align}
\end{subequations}
where (\ref{equ:optimal TDD constraint for jointly considering uplink and downlink in Optimum system a}) and (\ref{equ:optimal TDD constraint for jointly considering uplink and downlink in Optimum system b}) reflect the constraints of energy harvesting capacity and information transmit capacity, respectively. And (\ref{equ:optimal TDD constraint for jointly considering uplink and downlink in Optimum system c}) means the non-negativity of $R_d$ and $R_u$.

\emph{\textbf{Proposition 5.1:} When average transmit power is $P_0$ and downlink information rate is $R_d$, the capacity-rate function $C(R_d)$ Under optimal transmission policy is}
\begin{equation}\label{equ:the optimal solution in TDD in upper case}
  C(R_d)= (1-\lambda^\ast)B \cdot \log [1+\tfrac{Gh}{(1-\lambda^\ast)d^\alpha \sigma_a^2} \cdot (\tfrac{P_0 \eta Gh}{d^\alpha}-P_p)]
\end{equation}
\emph{where $\lambda^\ast$ denotes optimal time slot allocation ratio for downlink ($W(\cdot)$ is Lambert W function).}
\begin{equation}\label{equ:the optimal ratio in TDD in upper case}
  \lambda^\ast=\tfrac{-1}{\frac{B}{R_d \ln 2} W(-\frac{R_d \ln 2}{A_0 B} \cdot e^{-\frac{R_d}{A_0B}\ln 2}) +\tfrac{1}{A_0}}, \,\, A_0=\tfrac{GhP_0}{d^\alpha \sigma_a^2}
\end{equation}
\begin{proof}
 See Appendix C.
\end{proof}

\subsection{Optimal Transmission Policy in the Orthogonal Receiver System}

In orthogonal receiver system, assuming the power ratio allocated for energy sub-receiver is $\rho$, the relationship between downlink information rate and harvested energy power is given in Eqn. (\ref{equ:achievable capacity-energy region for R-d and P-s in TPSS}). Similar to the solution in the previous subsection, the optimization problem for $C(R_d)$ in this case should be rewritten as
\begin{subequations}\label{equ:optimal TDD for jointly considering uplink and downlink}
\begin{align}
 \max \limits_{\lambda,\rho} \,\,\,\,&  R_u = (1-\lambda) B\log(1+\tfrac{Gh P_u/(1-\lambda)}{ d^\alpha \sigma_0^2})  \tag{23}\\
   s.t.\,\,\,\,\,\,\, &   P_p + P_u \leq \lambda \eta \tfrac{Gh}{\lambda d^\alpha} \rho P_0  ;\,\,\,\,\,\,\,\,\,\,\,\,\,\,\,\,\,\,\,\,\,\,\,\,\, \label{equ:optimal TDD constraint for jointly considering uplink and downlink a} \\
   & R_d \leq \lambda B \log(1+\tfrac{Gh (1-\rho)P_0/\lambda}{ d^\alpha \sigma_0^2}); \label{equ:optimal TDD constraint for jointly considering uplink and downlink b} \\
   & R_d,R_u \geq 0. \label{equ:optimal TDD constraint for jointly considering uplink and downlink c}
\end{align}
\end{subequations}

We solve the problem in (\ref{equ:optimal TDD for jointly considering uplink and downlink}) from another way. When downlink and uplink information transmission rates are $(R_d,R_u)$, the average transmit power at master node, denoted as $P(\lambda)$, can be expressed as a function with respect to $\lambda$ on the condition of $R_d$ and $R_u$.
\begin{equation}\label{equ:expression of transmit power}
  P(\lambda)= \tfrac{d^\alpha}{\eta Gh} \cdot \{ \tfrac{(1-\lambda)d^\alpha \sigma_0^2}{Gh}[2^{\tfrac{R_u}{(1-\lambda)B}}-1]+ P_p\} +\tfrac{\lambda d^\alpha \sigma_0^2}{Gh} \cdot (2^{\tfrac{R_d}{\lambda B}}-1)
\end{equation}

\emph{\textbf{Proposition 5.2:} The function $P(\lambda)$ as shown in Eqn. (\ref{equ:expression of transmit power}) is a convex function with respect to variable $\lambda$ on the condition of $R_d$ and $R_u$.}
\begin{proof}
 See Appendix D.
\end{proof}

The residual problem is to find the maximum value of $R_u$ by adjusting $\lambda$ when $P(\lambda)=P_0$ and $R_d$ is fixed. Based on \emph{Proposition 5.2}, some heuristic iterative algorithm can be employed here to find the optimal value of $\lambda$ that corresponds to the maximum value of $R_u$, namely $C(R_d)$. Obviously, the minimum possible value of $R_u$ is $R_{u,min}=0$ while maximum possible value $R_{u,max}$ corresponds to the case $R_d=0$. The iterative algorithm is presented as \emph{Algorithm 1}.

\begin{algorithm}[htb]
\caption{Calculate $\lambda^\ast$ and $C(R_d)=\max \limits_{\lambda} \{R_u(\lambda)\}$}
\label{alg:Framwork}
\begin{algorithmic}[1] 
\REQUIRE ~~\\ 
The average transmit power $P_0$, downlink information rate $R_d$ and the range of $R_u$ value $[R_{u,min},R_{u,max}]$;\\
\ENSURE ~~\\ 
The optimal value of $\lambda^\ast$ and the maximum value of $R_u$, namely $C(R_d)$;
\STATE $R_u=(R_{u,min}+R_{u,max})/2$;\label{algorithem_begin}
\STATE For the given $R_d$ and $R_u$, finding the value $\lambda^\ast$ that minimizes the value of objective function $P(\lambda)$ in Eqn. (\ref{equ:expression of transmit power}) by convex optimization tools.
\IF{$P(\lambda^\ast)>P_0$}
\STATE $R_{u,max}=R_u$;
\ELSE
\STATE $R_{u,min}=R_u$;
\ENDIF
\IF{$|P(\lambda^\ast)-P_0|>0.001$}
\STATE return to Step (\ref{algorithem_begin});
\ELSE
\STATE goto to Step (\ref{algorithm_end});
\ENDIF
\STATE output $\lambda^\ast$ and $C(R_d)=R_u$;\label{algorithm_end}
\end{algorithmic}
\end{algorithm}

\subsection{Optimal Transmission Policy in the Power Splitting Receiver System}

In this subsection, let's consider the optimal policy for downlink and uplink information transmission in power splitting receiver system. The relationship between harvested energy and downlink information rate is given in Eqn. (\ref{equ:achievable capacity-energy region for R-d and P-s in PSRS}). When the uplink information rate $R_u$ is given, the optimization problem for capacity-rate function $C(R_u)$ can be modeled as follows.
\begin{subequations}\label{equ:optimal TDD for jointly considering uplink and downlink in PSRS}
\begin{align}
 \max \limits_{\lambda,\rho} \,\,\,\,& R_d \leq \lambda B \log(1+\tfrac{Gh (1-\rho)P_0/\lambda}{(1-\rho) \sigma_a^2+\sigma_p^2})   \tag{25}\\
   s.t.\,\,\,\,\,\,\, &   P_p + P_u \leq \lambda \eta \tfrac{Gh}{\lambda d^\alpha} \rho P_0  ;\,\,\,\,\,\,\,\,\,\,\,\,\,\,\,\,\,\,\,\,\,\,\,\,\, \label{equ:optimal TDD constraint for jointly considering uplink and downlink in PSRS a} \\
   & R_u = (1-\lambda) B\log(1+\tfrac{Gh P_u/(1-\lambda)}{ d^\alpha \sigma_0^2}); \label{equ:optimal TDD constraint for jointly considering uplink and downlink in PSRS b} \\
   & R_d,R_u \geq 0. \label{equ:optimal TDD constraint for jointly considering uplink and downlink in PSRS c}
\end{align}
\end{subequations}

Due to the monotonicity, in order to maximize $R_d$, the right side term should be equal to the left side term in the constraints (\ref{equ:optimal TDD constraint for jointly considering uplink and downlink in PSRS a}) and (\ref{equ:optimal TDD constraint for jointly considering uplink and downlink in PSRS b}).
By eliminating $\rho$, the capacity-rate function in this case can be rewritten as:
\begin{equation}\label{equ:rewritten the problem under optimal strategy}
  C(R_u)=\max \limits_{\lambda} \,\,\,\{ \lambda B \log(1+\tfrac{Gh (1-\rho(\lambda))P_0/\lambda}{(1-\rho(\lambda)) \sigma_a^2+\sigma_p^2})\}
\end{equation}
where
\begin{equation}\label{equ:expression of g-ratio}
  \rho(\lambda)=\tfrac{d^\alpha}{\eta GhP_0} \cdot (\tfrac{(1-\lambda)d^\alpha \sigma_0^2}{Gh}(2^{\frac{R_u}{(1-\lambda)B}}-1)+ P_p)
\end{equation}

With the help of $C(R_u)$, the achievable capacity-rate region in power splitting receiver system can be expressed as
\begin{equation}\label{equ:achievable capacity-rate region in power splitting receiver system}
  \mathbb{T}_{\{R_d-R_u\}}=\{(R_d,R_u)|0 \leq R_d \leq C(R_u),0 \leq R_u \leq R_{u,max}\}
\end{equation}

The residual task is to find the optimal $\lambda^\ast$ within the range $\lambda \in [0,1]$ that maximizes the value of $R_d$, namely $C(R_u)$. As observed from Eqn. (\ref{equ:rewritten the problem under optimal strategy}), $R_d$ can be expressed as a function of $\lambda$, denoted $R_d(\lambda)$. Though an explicit solution can't be found, simulated annealing algorithm is a good choice to calculate a sub-optimal $\lambda^\ast$ and $C(R_u)=R_d(\lambda^\ast)$ with acceptable computation complexity, which can be viewed as a lower bound system performance for power splitting receiver system. The calculation processing for $\lambda^\ast$ can be summarized as \emph{Algorithm 2}.

\begin{algorithm}
\caption{Calculate  $\lambda^\ast$ and $C(R_u)=\max \limits_{\lambda} \{R_d(\lambda)\}$}
\label{algorithm-1}
\begin{algorithmic}[1]
\REQUIRE $T,T_{min},r(0<r<1)$
\ENSURE $\lambda_{i},R_d(\lambda_{i})$
\STATE $\mathbf{Initialization:}$
\STATE $i=1,\lambda_{1}=random(0,1);$
\WHILE{$T>T_{min}$}
\STATE $\lambda_{i+1}=random(0,1);$
\STATE $dE=R_d(\lambda_{i+1})-R_d(\lambda_{i});$
\IF{$dE \geq 0$}
\STATE accept the mobility $\lambda_{i+1}$;
\ELSE
\IF{$e^{dE/T}<random(0,1)$}
\STATE $\lambda_{i+1}=\lambda_{i}$, don't accept the mobility;
\ENDIF
\ENDIF
\STATE $T=r*T;$
\STATE $i=i+1;$
\ENDWHILE
\end{algorithmic}
\end{algorithm}

\subsection{Simulation Results and Discussions}

Some simulation results will be given here to validate our results. Assuming the distance between master node and child node $d$ is $50 m$, the antenna power gain $G$ is $10 dB$, channel power gain coefficient $h$ is $1$, path loss component $\alpha$ is 2 and the frequency bandwidth $B$ is $1$ KHz. The overall additive noise $\sigma_0^2$ is $1 mW$. The basic power consumption for signal processing component $P_p$ is $5 mW$. The average transmit power at master node is $10 W$.

\begin{figure}[!t]
\centering
\includegraphics[width=3.4 in]{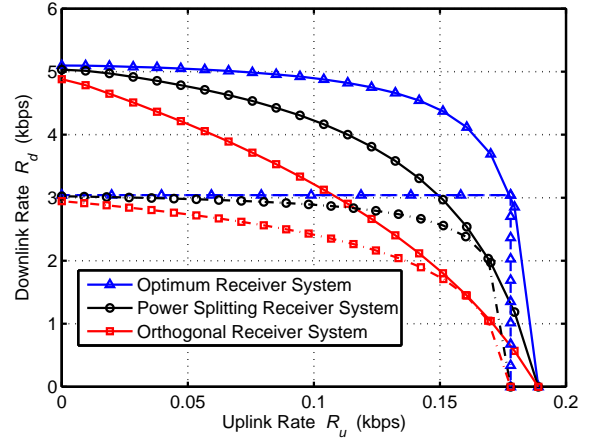}
\caption{The achievable capacity-rate region $\mathbb{T}_{\{R_d-R_u\}}$ which contains three different receiver structures: optimum receiver system, orthogonal receiver system and power splitting receiver system with $\sigma_p^2/\sigma_a^2=1/4$. The solid line represents optimal time allocation strategy is employed while dashed line represents halving time allocation strategy.}
\label{fig:result_three_case_sum}
\end{figure}

\begin{figure}[!t]
\centering
\includegraphics[width=3.2 in]{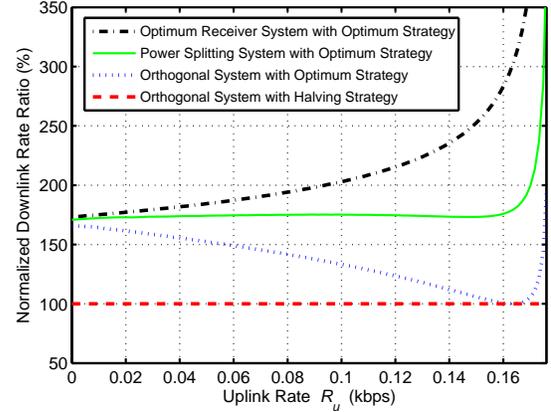}
\caption{The performance of different systems in terms of normalized downlink rate ratio as the uplink rate increasing from $0$ to $R_{u,max}$, where orthogonal system with halving allocation strategy is served as baseline.}
\label{fig:result_for_two_rate_region_normalized}
\end{figure}

Firstly, we consider the achievable capacity-rate region $\mathbb{T}_{\{R_d-R_u\}}$ under optimal time allocation strategy, which contains three different systems: optimum receiver system, orthogonal receiver system and power splitting receiver system with $\sigma_p^2/\sigma_a^2=1/4$. A simple time allocation strategy for TDD system that halves the time resource for uplink and downlink is also given, to act as a comparison. The two kinds of systems are called as optimal strategy system and halving strategy system in this subsection, respectively.
Fig. \ref{fig:result_three_case_sum} illustrates the performance curve of energy harvesting system in terms of $\mathbb{T}_{\{R_d-R_u\}}$. The solid line represents optimal time allocation strategy is employed while dashed line represents halving time allocation strategy. Obviously, optimal strategy system is superior to halving strategy system, which benefits from the optimal time resource allocation.
Fig. \ref{fig:result_for_two_rate_region_normalized} compares the performance of different systems in terms of normalized downlink rate ratio as the uplink rate increasing from $0$ to $R_{u,max}$, where orthogonal system with halving allocation strategy is served as baseline. The other three cases in the figure are orthogonal system, power splitting system and optimum receiver system with optimum transmission strategy.
It can be seen that substantial benefit can be obtained from optimum receiver structure and optimum allocation strategy, the value of which is bigger than $70\%$ in most case.

\begin{figure}[!t]
\centering
\includegraphics[width=3.2 in]{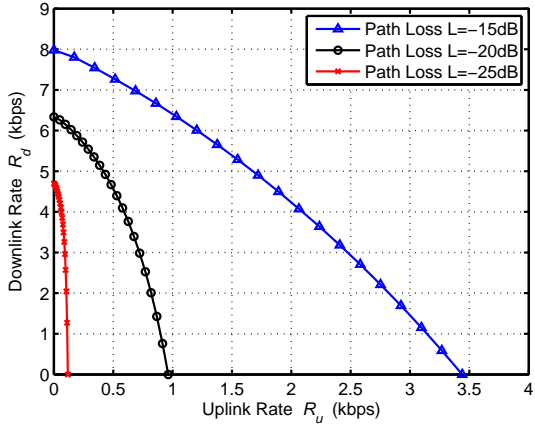}
\caption{The achievable capacity-rate region $\mathbb{T}_{\{R_d-R_u\}}$ under different path loss values when transmit power $P_0$ is $10 W$ in power splitting receiver system with $\sigma_p^2/\sigma_a^2=1/4$.}
\label{fig:result_different_gain_general}
\end{figure}

As observed from Eqn. (\ref{equ:expression of transmit power}), in order to transmit unit information, the required transmit power for uplink is nearly twice as much as that for downlink in terms of $dB$. That is to say, the power used to support uplink rate experiences higher path loss since it experiences longer transmission distance. Path loss $L$ is denoted as $L=G/d^\alpha$. Fig. \ref{fig:result_different_gain_general} illustrates the achievable capacity-rate region $\mathbb{T}_{\{R_d-R_u\}}$ under different path loss values when transmit power $P_0$ is $10 W$ in power splitting receiver system with $\sigma_p^2/\sigma_a^2=1/4$. It can be seen that $R_u$ decreases faster than $R_d$ as the increment of path loss $L$, which is similar to the \emph{doubly near-far} phenomenon proposed in \cite{Ju_32}. As a consequence, if the distance $d$ between transmitter and receiver is big, the power efficiency in RF-based energy harvesting system will become unacceptable. A feasible solution is to increase the power gain factor $G$ generated by antenna, such as beam-forming technology.

In some practical system, in addition to transmit and receive signal, there are some other usages that need energy supply at child node. For example, the node in sensor network also needs energy to keep awake and collect data. Assuming the average residual power that is needed for these usages is $P_r$, the corresponding performance metric is \emph{achievable capacity-energy-rate $\mathbb{T}_{\{R_d-P_r-R_u\}}$}. By modifying the constraint in (\ref{equ:optimal TDD constraint for jointly considering uplink and downlink in PSRS a}) into (\ref{equ:new constraint for three dimension optimal TDD}), we can use the model in (\ref{equ:optimal TDD for jointly considering uplink and downlink in PSRS}) to obtain the solution to $\mathbb{T}_{\{R_d-P_r-R_u\}}$.
\begin{equation}\label{equ:new constraint for three dimension optimal TDD}
 P_r+P_p+P_u \leq \lambda \eta \tfrac{Gh}{\lambda d^\alpha} \rho
\end{equation}

\begin{figure}[!t]
\centering
\includegraphics[width=3.2 in]{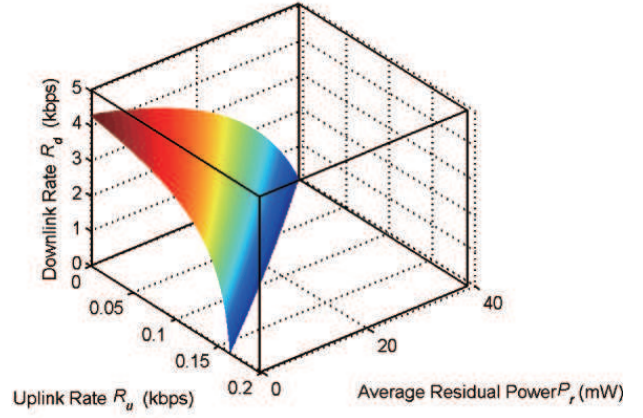}
\caption{The achievable capacity-energy-rate $\mathbb{T}_{\{R_d-P_r-R_u\}}$ When the average transmit power is $10 W$ in power splitting receiver system with $\sigma_p^2/\sigma_a^2=1/4$.}
\label{fig:result_three_dimension}
\end{figure}

The solution for solving $\mathbb{T}_{\{R_d-P_r-R_u\}}$ is similar to the solution to problems in (\ref{equ:optimal TDD for jointly considering uplink and downlink in Optimum system}), (\ref{equ:optimal TDD for jointly considering uplink and downlink}) and (\ref{equ:optimal TDD for jointly considering uplink and downlink in PSRS}), which is neglected in this paper due to limit of space. Taking the power splitting receiver system with $\sigma_p^2/\sigma_a^2=1/4$ for example, when the average transmit power is $10 W$ and other parameters are just the same as before, the corresponding $\mathbb{T}_{\{R_d-P_r-R_u\}}$ is shown in Fig. \ref{fig:result_three_dimension}, which indicates the tradeoff relationship between two information rates and residual energy $P_s$. The result shown in Fig. \ref{fig:result_three_case_sum} can be regarded as an extreme case of $\mathbb{T}_{\{R_d-P_r-R_u\}}$ when $P_s$ is $0$.

\section{A Way to Green System}

The fundamental limits of information transmission in wireless communication system with RF-based energy harvesting has been investigated. Due to huge path loss, energy harvested from RF signal is so rare and precious that it is very important to employ appropriate strategies to improve the overall energy efficiency. Thus, for giving some more insights for energy harvesting system design from a perspective of system-level, a typical application in terms of minimizing required transmit power, namely green system, will be introduced in the sequel.

Considering a data acquisition system in body-area network whose structure is shown in Fig. \ref{fig:system_framework}(a). The child node may be a sensor embedded inside human body while external master node is the information and control centre for whole system. Master node transmits control information and energy to child node, thus child node isn't constrained by the battery's limited-lifetime. Then, child node uses the harvested energy to transmit the acquired data back to master node.
Assuming the distance between master node and child node $d$ is $2 m$, the antenna power gain $G$ is $0 dB$, channel power gain coefficient $h$ is $1$, path loss component $\alpha$ is 2 and the frequency bandwidth $B$ is $1$ KHz. The additive channel noise $\sigma_0^2$ is $1 mW$ and  signal processing power consumption at child node $P_p$ is $5 mW$.

\begin{figure}[!t]
\centering
\includegraphics[width=3.2 in]{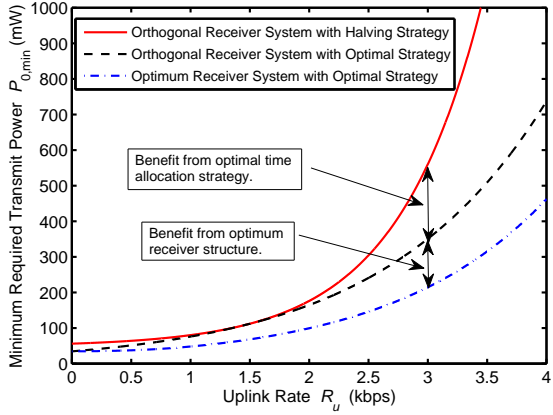}
\caption{The relationship between minimum required transmit power $P_{0,min}$ and uplink rate $R_u$ when $R_d=2 kbps$.}
\label{fig:application_continue_service}
\end{figure}

It is assumed that the expected average information rate from master node to child node is $R_d=2 kbps$. Fig. \ref{fig:application_continue_service} illustrates the value of minimum required transmit power $P_{0,min}$ as uplink rate $R_u$ increased from 0 to $4 kbps$, which contains three cases: orthogonal receiver system with halving (optimal) time allocation strategy and optimum receiver system with optimal time allocation strategy. From Fig. \ref{fig:application_continue_service}, when $R_d=2 kbps$ and $R_u=3 kbps$, the benefit from optimal time allocation strategy is about saving $37\%$ transmit power compared to halving strategy system. What' more, if optimum receiver is applied, more $39\%$ transmit power can be saved compared to orthogonal receiver system with optimal time allocation strategy, which is the extreme benefit limit bringed from state-of-art.

\section{Conclusion}

This paper has investigated the transmission performance of wireless communication system with RF-based energy harvesting, which consists of a master node and a child node. Child node is powered by the energy harvested from master node. From a perspective of information transmission, two way information rates between master node and child node are the ultimate indicators of system performance under an unique external power supply at master node.

As a result, this paper firstly reviewed the problem for simultaneous information/energy transfer in three typical system structures, namely optimum receiver system, orthogonal receiver system and power splitting system. Then, the harvested energy managing strategy and transmission capacity from child to master node were analyzed.
After that, we jointly considered the two way information transmission between two nodes from a view of systemic level in a time division duplex system. And power consumption by signal processing at child node is also taken into account throughout this paper. For characterizing the tradeoff relationship between two way information rates, \emph{achievable capacity-rate region} is proposed by this paper to use as a metric to evaluate transmission performance.

By formulating constrained optimization problem, the boundary expression of \emph{achievable capacity-rate region}, namely the most energy-efficient status, is derived with the help of \emph{conditional capacity function}. One conclusion that can be drawn from simulation results is that substantial benefits can be obtained from employing optimum receiver system and optimal transmission strategy in terms of green communication. According to the results developed by this paper, an typical example in body-area network that employs energy harvesting technology was introduced from a perspective of minimizing the required transmit power. Three different related systems were compared with each other by simulation results. Besides, the extreme benefits that can be obtained from optimal transmission strategy and optimum receiver structure were also indicated by simulation results in terms of minimum required transmit power.



\section{Appendices}
\subsection{Proof of the Proposition 3.1}

It is assumed that the average harvested energy power and information transmission rate are denoted as $P_h$ and $R_d$, respectively. When the ratio of transmit power allocated for energy sub-receiver is $\rho$ and the time ratio allocated for wireless energy transfer is $\tau$, the instantaneous transmit power for energy sub-receiver during transmission is $\rho P_0/\tau$. According to wireless energy transfer capacity in Eqn. (\ref{equ:model for energy receiver}), $P_h$ can be expressed as
\begin{equation}\label{equ:harvested energy expression in TS}
  P_{h,max}=\tau \eta \frac{Gh}{d^\alpha} \cdot \frac{\rho P_0}{\tau}= \eta \frac{Gh\rho P_0}{d^\alpha}
\end{equation}

Similarly, the information transmission rate $R_d$ under above consumption can be expressed as
\begin{equation}\label{equ:information rate expression in TS}
  R_d=\tau \log (1+\frac{Gh}{\sigma_a^2+\sigma_p^2}\cdot \frac{(1-\rho)P_0}{\tau})
\end{equation}

From (\ref{equ:harvested energy expression in TS}) and (\ref{equ:information rate expression in TS}), it is obvious that $P_h$ is independent with $\tau$ while $R_d$ is monotone-decreasing function with $\tau\in [0,1]$. In terms of maximizing overall performance, $\tau$ should be as small as possible. The only limit is the maximum peak power of transmitter. For example, if maximum peak power at transmitter is $P_{max}$, the instantaneous transmit power for energy sub-receiver should satisfy: $\rho P_0/\tau \leq P_{max}$. And the minimum achievable value of $\tau$ is $\rho P_0/P_{max}$.

As a result, if we consider the ideal case that $P_{max}\rightarrow \infty$, the optimal value of $\tau$ is asymptotic to zero and instantaneous power is $\rho P_0 \delta(\tau)$. In this case, the achievable capacity-energy region based on the \emph{Definition 3.1} can be expressed as
\begin{equation}\label{equ:achievable capacity-energy region for R-d and P-s in appendix A}
\begin{split}
  & \mathbb{T}_{\{R_d-P_s\}} =\,\bigcup \limits_{0 \leq \rho \leq 1} \,\{\,(R_d,P_s)\,\,|\,0 \,\leq \,P_s+P_p \\
  & \leq \eta \tfrac{\rho Gh}{d^\alpha} P_0, 0\, \leq \, R_d \,\leq \, B\log(1+\tfrac{Gh (1-\rho)P_0}{d^\alpha \sigma_0^2})\}
\end{split}
\end{equation}

Thus, \emph{Proposition 3.1} has been proved.


\subsection{Proof of the Proposition 4.1}

Let's solve the problem in (\ref{equ:optimal problem under random arrival energy}) from a dual perspective in terms of minimizing energy consumption. It is assumed that data rate requirement in $n-th$ time slot is $C(n)$, and the minimum power requirement for $C(n)$ is denoted as $P_c(n)$. Based on Eqn. (\ref{equ:optimal problem constraint under random arrival energy a}), $P_c(n)$ can be expressed as
\begin{equation}\label{eqn:the expression of power consumption}
  P_c(n)=\frac{d^\alpha\sigma_0^2}{Gh}(2^{\frac{C(n)}{B}}-1)+P_p
\end{equation}

It can be seen from Eqn. (\ref{eqn:the expression of power consumption}) that $P_c(n)$ is a convex function with respect to $C(n)$. As a result, transmit signal with a constant rate is optimal in terms of minimizing energy consumption. Combining the result in \emph{Lemma 4.1}, it can be concluded that the maximal achievable uplink information rate $R_u$ is expressed as Eqn. (\ref{equ:result of Proposition 1}).

Thus, \emph{Proposition 4.1} has been proved.

\subsection{Proof of the Proposition 5.1}

The problem is equivalent to solve following equation set with unknown $\lambda^\ast$ and $C(R_d)$:
\begin{subequations}\label{equ:equivalent problem for optimum receiver system}
\begin{align}
 & \frac{\lambda^\ast d^\alpha (\sigma_a^2+\sigma_p^2)}{Gh} \cdot (2^{\frac{R_d}{\lambda^\ast B}}-1)=P_0 \label{equ:equivalent problem for optimum receiver system a}   \\
 & \frac{d^\alpha}{\eta Gh} \cdot \{ \frac{(1-\lambda^\ast)d^\alpha \sigma_a^2}{Gh}[2^{\frac{C(R_d)}{(1-\lambda^\ast)B}}-1]
   + P_p\}=P_0  \label{equ:equivalent problem for optimum receiver system b}
\end{align}
\end{subequations}

Assuming $x=\frac{1}{\lambda^\ast}$ and $A_0=\frac{GhP_0}{d^\alpha (\sigma_a^2+\sigma_p^2)}$, Eqn. (\ref{equ:equivalent problem for optimum receiver system a}) can be transformed into:
\begin{equation}\label{equ:transformed equation for optimal allocation a}
  2^{\frac{R_d}{B}x}=A_0x+1
\end{equation}
Then
\begin{equation}\label{equ:transformed equation for optimal allocation c}
  -\tfrac{R_d}{A_0B}\ln 2  (A_0x+1)\cdot e^{-\frac{R_d}{A_0B}\ln 2 (A_0x+1)}=
             -\tfrac{R_d}{A_0B}\ln 2 \cdot e^{-\frac{R_d}{A_0B}\ln 2}
\end{equation}
By Lambert $W(\cdot)$ function, we can obtain:
\begin{equation}\label{equ:transformed equation for optimal allocation d}
  x=-\tfrac{B}{R_d \ln 2} W(-\tfrac{R_d}{A_0B}\ln 2 \cdot e^{-\frac{R_d}{A_0B}\ln 2}) -\tfrac{1}{A_0}
\end{equation}
Thus
\begin{equation}\label{equ:result 1 in appendix B}
  \lambda^\ast=\frac{1}{-\frac{B}{R_d \ln 2} W(-\frac{R_d}{A_0B}\ln 2 \cdot e^{-\frac{R_d}{A_0B}\ln 2}) -\frac{1}{A_0}}
\end{equation}

Substituting (\ref{equ:result 1 in appendix B}) into Eqn. (\ref{equ:equivalent problem for optimum receiver system b}), we can obtain:
\begin{equation}\label{equ:result 2 in appendix B}
  C(R_d)= (1-\lambda^\ast)B \cdot \log [1+\tfrac{Gh}{(1-\lambda^\ast)d^\alpha \sigma_a^2} \cdot (\tfrac{P_0 \eta Gh}{d^\alpha}-P_p)]
\end{equation}

\subsection{Proof of the Proposition 5.2}

Assuming $f(\lambda)=A_0\lambda \cdot (e^{\frac{B_0}{\lambda}}-1)$, where $0 \leq \lambda \leq 1$ and $A_0,B_0 >0$. Then the second-order derivative of $f(\lambda)$ is
\begin{equation}\label{eqn:second-order derivative}
  \tfrac{\partial^2 f(\lambda)}{\partial \lambda^2}=\tfrac{A_0B_0^2}{\lambda^3}\cdot e^{\frac{B_0}{\lambda}} >0
\end{equation}

Obviously, $f(\lambda)$ is convex with respect to $\lambda$. Let $A_0=\tfrac{d^\alpha \sigma_0^2}{Gh}$ and $B_0=\tfrac{R_d}{B}$, it can be obtained that the function $f_1=\tfrac{\lambda d^\alpha \sigma_0^2}{Gh} \cdot (2^{\tfrac{R_d}{\lambda B}}-1)$ is convex with respect to $\lambda$.

Similarly, it can be proved that $f_2=\tfrac{(1-\lambda)d^\alpha \sigma_0^2}{Gh}[2^{\tfrac{R_u}{(1-\lambda)B}}-1]$ is also convex with respect to $\lambda$. Since the sum of two convex functions and a linear function is still a convex function, $P(\lambda)$ is a convex function with respect to variable $\lambda$ on the condition of $R_d$ and $R_u$.

Thus, \emph{Proposition 5.2} has been proved.



\ifCLASSOPTIONcaptionsoff
  \newpage
\fi

\end{document}